\documentclass[%
 aip,
 amsmath,amssymb,
 reprint,%
]{revtex4-1}
\usepackage{graphicx}% Include figure files
\usepackage{dcolumn}% Align table columns on decimal point
\usepackage{bm}% bold math
\usepackage{xcolor}
\usepackage{comment}
\usepackage[utf8]{inputenc}
\usepackage[T1]{fontenc}
\usepackage{mathptmx}
\usepackage{etoolbox}
\makeatletter
\def\@email#1#2{%
 \endgroup
 \patchcmd{\titleblock@produce}
  {\frontmatter@RRAPformat}
  {\frontmatter@RRAPformat{\produce@RRAP{*#1\href{mailto:#2}{#2}}}\frontmatter@RRAPformat}
  {}{}
}%
\makeatother
\begin{document}

\preprint{AIP/123-QED}

\title[Multiplexing-based control of wavefront propagation]{Multiplexing-based control of wavefront propagation: the interplay of inter-layer coupling, asymmetry and noise}
% Force line breaks with \\
\author{Vladimir V. Semenov}
\email{semenov.v.v.ssu@gmail.com}
\affiliation{Institut f\"{u}r Theoretische Physik, Technische Universit\"{a}t Berlin, Hardenbergstra{\ss}e 36, 10623 Berlin, Germany}
\affiliation{Institute of Physics, Saratov State University, 83 Astrakhanskaya str., 410012 Saratov, Russia}

\author{Sarika Jalan}
\affiliation{Complex Systems Lab, Department of Physics, Indian Institute of Technology Indore, Khandwa Road, Simrol, Indore-453552, India}

\author{Anna Zakharova}
\affiliation{Institut f\"{u}r Theoretische Physik, Technische Universit\"{a}t Berlin, Hardenbergstra{\ss}e 36, 10623 Berlin, Germany}
\affiliation{Bernstein Center for Computational Neuroscience, Humboldt-Universit\"{a}t zu Berlin, Philippstra{\ss}e 13, 10115 Berlin, Germany}
\date{\today}% It is always \today, today,

\begin{abstract}
We show how multiplexing influences propagating fronts in multilayer networks of coupled bistable oscillators. Using numerical simulation, we investigate both deterministic and noise-sustained propagation. In particular, we demonstrate that the multiplexing allows to reduce the intra-layer dynamics to a common regime where the front propagation speed in all the interacting layers attains the same fixed value. In the presence of noise the dynamics is more complicated and is characterized by the ability of the system to adjust to the common propagation speed for varying the multiplexing strength. In addition, we find that the noise-induced stabilization of wavefront propagation in multilayer networks allows to obtain less pronounced deviations of the wavefront compared to the stabilization achieved in the isolated layer. Finally, we demonstrate that the reduction of the wavefront deviations can be enhanced by increasing the number of interacting layers. 
\end{abstract}

%\item In systems with complex dynamics (lumped and distributed) interacting by means of inertial nonlinear coupling, a regime of complete synchronization is observed
%\item In the case of connection through memristors, the synchronization threshold significantly depends on the initial state of the coupling elements
%\item The dependence of the synchronization threshold on the initial state of the coupling elements is observed not only in the case of connection through ideal memristors, but also in the case of non-ideal memristors, which are nonlinear elements with strong inertiality

\maketitle

\section{\label{sec:intro}Introduction}
A broad variety of spatially-extended dynamical systems can experience a non-equilibrium transition manifested though formation and propagation of domains and waves  \cite{mikhailov1990,kapral1995,garcia-ojalvo1999}. The simplest types of dynamical systems exhibiting such propagation are ensembles and media, where bistability results from the coexistence of two steady states in the phase space of individual oscillators. In such a case, a system evolves from its initial state such that two kinds of domains corresponding to quiescent steady state regimes are formed in space. After that, the domains as well as the boundaries between them called 'fronts' or 'wavefronts'  can propagate. The wavefront propagation in bistable media plays an important role in terms of chemistry (for instance, see Schl\"{o}gl model developed to describe an autocatalytic reaction mechanism \cite{schloegl1972,schloegl1983,loeber2014}), flame propagation theory \cite{zeldovich1938}, electronics \cite{schoell2001}, to name just a few.
Domain growth and wavefront propagation observed in 2D- and 3D-space is often referred to as 'coarsening'. It occurs in the same way when compared to classical front propagation in 1D-space, but with respect to the shape of domains. Coarsening represents a fundamental phenomenon and unites a wide spectrum of processes studied in the context of physics of liquid crystals \cite{yurke1992} and magnetism \cite{bray1994,cugliandolo2010,caccioli2008,denholm2019}, physics and chemistry of materials \cite{goh2002,zhang2019,zhang2019-2,geslin2019}, laser physics \cite{yanchuk2012,marino2014,javaloyes2015}, electronics \cite{semenov2018} and animal population statistics \cite{dobramysl2018}. Besides spatially-extended systems \cite{bray1994}, coarsening can be exhibited by single time-delay oscillators prepared in an inhomogeneous state characterized by coexisting equilibria. This effect becomes clearly identified when purely temporal dynamics of time-delay oscillators is mapped on space by means of virtual space-time representation \cite{yanchuk2012,marino2014,semenov2018}. In addition to the occurrence in deterministic systems, propagating fronts and coarsening processes can emerge in stochastic systems as phenomena accompanying the noise-induced phase transitions \cite{vandenbroeck1997,carrillo2003}.

The ability to demonstrate propagating fronts and coarsening is connected directly to the symmetry properties of bistable systems and with the spatial interaction intensity (for instance, diffusion coefficient in reaction-diffusion models or coupling strength in ensembles of interacting oscillators). To control the front propagation speed and direction, one can vary parameters responsible for the symmetry of the local dynamics as well as adjust the interaction strength \cite{loeber2014}. In addition, one can apply stochastic forcing for this purpose. In particular, it is known that multiplicative noise influences the front dynamics \cite{engel1985,garcia-ojalvo1999,mendez2011}. 

In the present work, we introduce a new scheme for controlling the front propagation, which can be implemented in multilayer networks of bistable oscillators. We show that connecting a one-layer network to another one-layer network through coupling between replica nodes, i.e., multiplexing, provides a tool for controlling the front propagation. Multiplexing-based schemes have been applied for controlling a wide spectrum of phenomena. For instance, such approach has been reported for deterministic networks with static inter-layer topology in the context of chimera \cite{ZAK20} and solitary \cite{SCH21} states as well as for controlling transitions between two network states \cite{MIK19,RYB21}. Moreover, it has been demonstrated that topological asymmetry in multilayer networks provides stabilization of chaotic dynamics manifested by the establishing of stable periodic orbits and equilibria (so-called asymmetry-induced order) \cite{MED21}. Furthermore, the significant impact of the dynamic inter-layer topology allows to achieve the inter-layer synchronization at lower inter-layer coupling strength \cite{ESE21}.

%%%%%%%%%%%%%%%%%%fig1%%%%%%%%%%%%%%%%%%%%%%
\begin{figure*}[t!]
\centering
\includegraphics[width=0.95\textwidth]{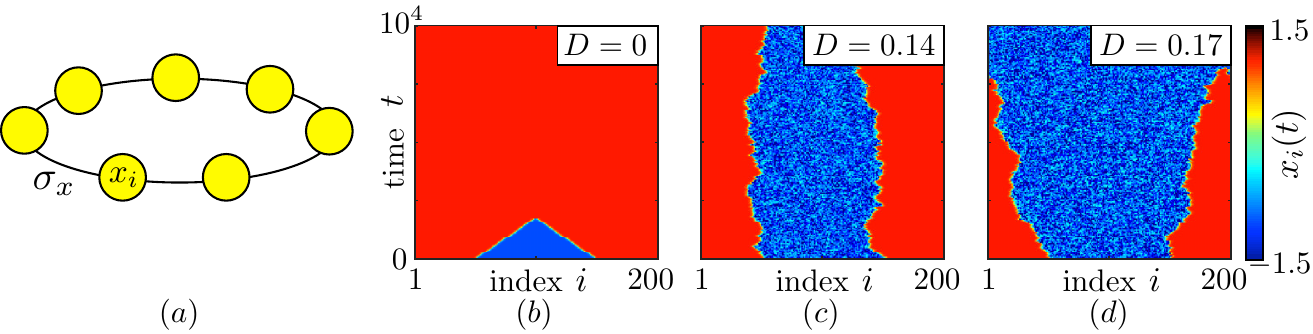}
\caption{Stochastic control of front propagation (coarsening) in a single-layer network (Eq. (\ref{single-layer}) in the presence of parametric noise in each oscillator: $b_x=0.9+\sqrt{2D}n_i(t)$). (a) Schematic representation of a one-layer network (layer $x$); (b)-(d) Spatio-temporal dynamics for increasing noise intensity, $D$. System parameters: $a=1$, $\sigma_x=1$.}
\label{fig1}
\end{figure*}  
%%%%%%%%%%%%%%%%%%%%%%%%%%%%%%%%%%%%%%%%%%

Multiplexing can have different impact on the dynamics in stochastic multilayer networks. In particular, multiplexing noise (noisy modulation of the inter-layer coupling strength) can be applied for the regulation of the inter-layer synchronization of spatio-temporal patterns in multilayer networks \cite{VAD20,RYB22}. A multiplexing-based approach \cite{semenova2018,MAS21,semenov2022} has been successfully applied to control the phenomenon of coherence resonance \cite{PIK97,MAS17,BAS21} observed in a multilayer networks of excitable oscillators as well as the phenomenon of stochastic resonance \cite{gammaitoni1998,anishchenko1999} exhibited by a multilayer network of bistable oscillators. For both cases, the multiplexing-based control allows to enhance or suppress the considered effects. Therefore, multiplexing can play both constructive and destructive role for the resonant stochastic phenomena associated with noise-induced regularity (coherence) of the stochastic dynamics. 

Here, we highlight a new facet of multiplexing in its impact on the deterministic and stochastic dynamics. The constructive role of multiplexing which we demonstrate here consists in two facts: varying the inter-layer coupling strength, one can (i) change the front propagation speed and direction, and (ii) minimize deviations of noise-driven fronts. The presented results can be potentially applied for controlling stochastic multilayer network dynamics actively studied in the context of deep learning \cite{semenova2019,semenova2022,semenova2022-2}. Indeed, the property of bistability can be easily achieved in various kinds of artificial neural networks including bistable neural networks with tanh-nonlinearity \cite{stern2014,vecoven2021}. Thus, one can expect the occurrence of front propagation and coarsening in such systems, which can potentially influence neural network characteristics and performance. For this reason, we expect that the presented results would be interesting for experts in artificial intelligence and machine learning besides specialists in nonlinear dynamics and theory of stochastic processes.

%%%%%%%%%%%%%%%%%%fig2%%%%%%%%%%%%%%%%%%%%%%
\begin{figure*}[t!]
\centering
\includegraphics[width=0.95\textwidth]{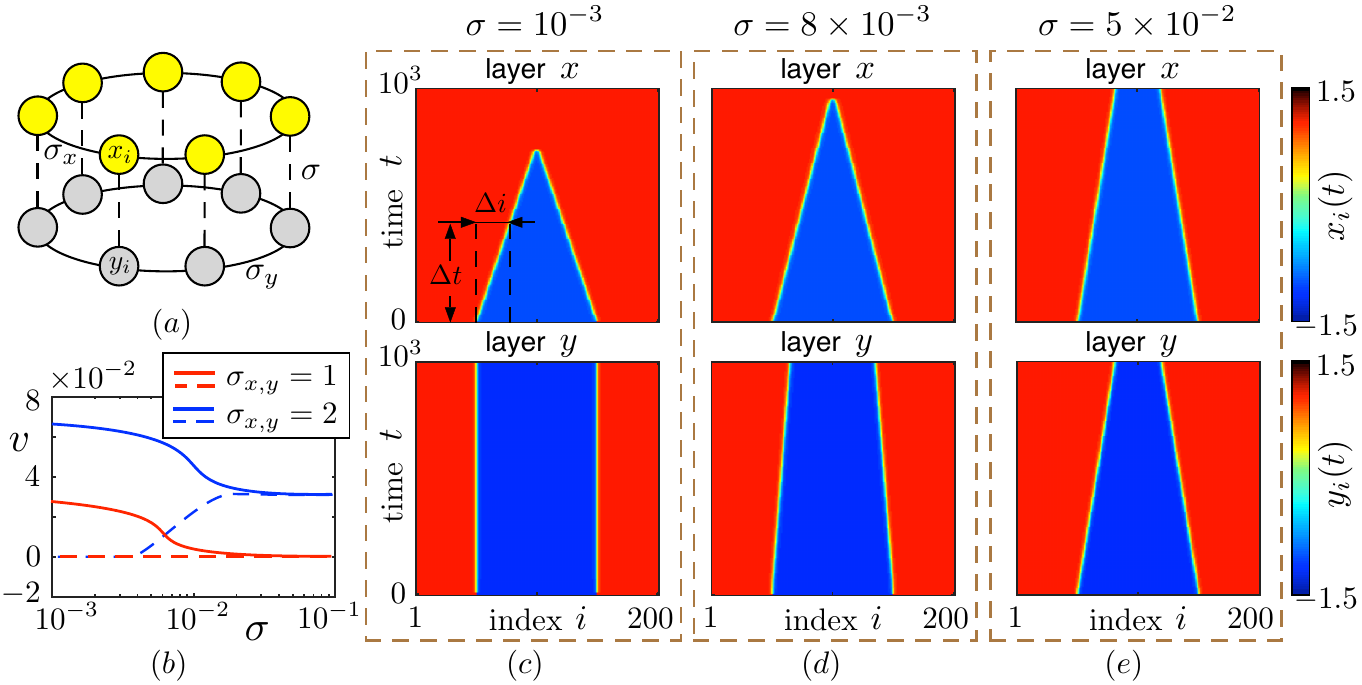}
\caption{Multiplexing-based control of front propagation (coarsening) in a two-layer multiplex network Eqs. (\ref{two-layers}) in the absence of noise. (a) Schematic representation of the network (layers $x$ and $y$); (b) Dependence of the front propagation speed on the multiplexing strength in layer $x$ (solid curves) and $y$ (dashed curves) for intra-layer coupling $\sigma_{x,y}=1$ (red curves) and $\sigma_{x,y}=2$ (blue curves); (c)-(e) Spatio-temporal dynamics of layer $x$ (upper panels) and $y$ (lower panes) for fixed  intra-layer coupling ($\sigma_x=\sigma_y=2$) and increasing multiplexing strength: $\sigma=10^{-3}$ (panels (c)), $\sigma=8\times10^{-3}$ (panels (d)), $\sigma=5\times10^{-2}$ (panels (e)). Other parameters: $a_x=1.0$, $b_x=0.9$, $a_y=1.0$, $b_y=1.0$.}
\label{fig2}
\end{figure*}  
%%%%%%%%%%%%%%%%%%%%%%%%%%%%%%%%%%%%%%%%%%

\section{\label{sec:single}Single-layer dynamics}
Before studying the impact of multiplexing, we discuss the case of a single ring of locally coupled overdamped bistable oscillators [Fig.~\ref{fig1}~(a)] in order to compare the isolated- and coupled-layer dynamics. The system equations take the following form:
\begin{equation}
\label{single-layer}
\begin{array}{l}
\dfrac{dx_{i}}{dt}=-x_{i}(x_{i}-a_x)(x_{i}+b_x)+\dfrac{\sigma_{x}}{2}\sum\limits^{i+1}_{j=i-1}(x_{j}-x_{i}), \\
%mx_{i}-x_{i}^3+A\sin(\omega_{\text{e}} t)+\sqrt{2D}n_i(t)\\
%+\dfrac{\sigma_{x}}{2}\sum\limits^{i+1}_{j=i-1}(x_{j}-x_{i}), \\
\end{array}
\end{equation}
where $x_i$ are the dynamical variables, $i=1, 2, ..., N$ with $N$ being the total number of elements in the layer. In this study, all the network layers consist of $N=200$ elements.  The strength of the coupling within the layer (intra-layer coupling) is given by $\sigma_{x}$. Parameters $a$ and $b$ determine the dynamics of individual network elements. They define whether the individual element nonlinearity is symmetric ($a_x=b_x$) or asymmetric ($a_x \neq b_x$). In the current study, we assume that all the elements are in the bistable regime ($a_x$ and  $b_x$ are positive). We perform our investigations by means of numerical simulations. In more detail, we integrate the differential equations numerically using the Heun method \cite{mannella2002} with the time step $\Delta t=0.001$ and the total integration time $t_{\text{total}}=10^4$. These numerical integration method parameters are chosen for modeling of all the cases discussed in the paper.

In fact, ensemble Eq. (\ref{single-layer}) represents the reaction-diffusion equation $\dfrac{dx}{dt}=-x(x-a_x)(x+b_x)+k\nabla^2 x$ rewritten for discretized space by means of the finite difference method (this technique is widely used for the modeling of reaction-diffusion systems \cite{strauss2007,linge2016}). It is well-known that bistable reaction-diffusion systems exhibit front propagation in the presence of asymmetry ($a \neq b$ in ensemble Eq. (\ref{single-layer})) \cite{loeber2014}. As mentioned in Sec. \ref{sec:intro}, the propagation can be stopped (stabilized) and then made inverse by using parametric noise. The model Eq. (\ref{single-layer}) demonstrates the same effect illustrated in Fig.\ref{fig1} (b)-(d) for the parameter values $a_x=1.0$, $b_x=0.9$, $\sigma_x=1.0$ and the initial conditions $x_i(t=0)=-b_x$ for $i\in [50:150]$ and $x_i(t=0)=a_x$ elsewhere (this kind of initial conditions is used throughout the paper). In the absence of noise, one observes expansion of the domain corresponding to the state $x_i(t)=a_x$  (red domain in Fig. \ref{fig1} (b)) which invades the entire available space. However, the front propagation can be slowed down, stopped [Fig. \ref{fig1} (c)] and reversed [Fig. \ref{fig1} (d)] in the presence of parametric noise modulating the parameter $b_x$ by increasing the noise level. Here, we introduce parametric noise modulating parameter $b_x$ of each oscillator $x_i$ in the form $b_x=0.9+\sqrt{2D}n_i(t)$. Further, $\sqrt{2D}n_i(t)\in\mathbb{R}$ is Gaussian white noise with intensity $D$, i.e., $<n_i(t)>=0$ and $<n_i(t)n_{j}(t)>=\delta_{ij}\delta(t-t')$, $\forall i,j$. The opposite effect is achieved if instead modulating parameter $b_x$, the parametric noise is introduced for the parameter $a_x$ in the form: $a_x=1.0+\sqrt{2D}n_i(t)$. In such a case, increasing noise speeds up the front propagation and the state $x_i=a_x$ invades the available space faster than in the deterministic case.

One aspect of the noise-based control is important to note here. It is shown in Fig. \ref{fig1} that one can control the mean front propagation speed by applying multiplicative noise and pass through the stabilized state corresponding to zero speed. However, the stochastic stabilization of the front propagation is achieved at high level of noise and the deviations of the instantaneous front position become evident and significant (see Fig. \ref{fig1} (c)). As we demonstrate in Sec. 3, such deviations can be minimized in a multilayer network due to the action of multiplexing. 

\section{\label{sec:multi}Multilayer network}
Next, we consider a two-layer multiplex network depicted in Fig.~\ref{fig2}~(a), where each layer represents a ring of locally coupled bistable oscillators. The oscillators  in the first layer $x_i$ are not under direct action of noise, while the second-layer oscillators $y_i$ contain statistically independent sources of multiplicative white Gaussian noise. The system equations take the form
\begin{equation}
\label{two-layers}
\begin{array}{l}
\dfrac{dx_{i}}{dt}=-x_{i}(x_{i}-a_x)(x_{i}+b_x)\\
+\dfrac{\sigma_{x}}{2}\sum\limits^{i+1}_{j=i-1}(x_{j}-x_{i})+\sigma(y_{i}-x_{i}), \\
\dfrac{dy_{i}}{dt}=-y_{i}(y_{i}-a_y)(y_{i}+b_y+\sqrt{2D}n_i(t))\\
+\dfrac{\sigma_{y}}{2}\sum\limits^{i+1}_{j=i-1}(y_{j}-y_{i})+\sigma(x_{i}-y_{i}).
\end{array}
\end{equation}
The strength of the coupling within the layer (intra-layer coupling) is given by $\sigma_{x}$ and $\sigma_{y}$ for the first and second layer, respectively. The coupling between the layers (inter-layer coupling) is bidirectional, diffusive and its strength is characterized by parameter $\sigma$. We consider a multiplex network, where the layers contain the same number of nodes and the inter-layer links are allowed only for replica nodes, i.e., there is a one-to-one correspondence between the nodes in different layers. 

%%%%%%%%%%%%%%%%%%fig3%%%%%%%%%%%%%%%%%%%%%%
\begin{figure*}[t]
\centering
\includegraphics[width=0.9\textwidth]{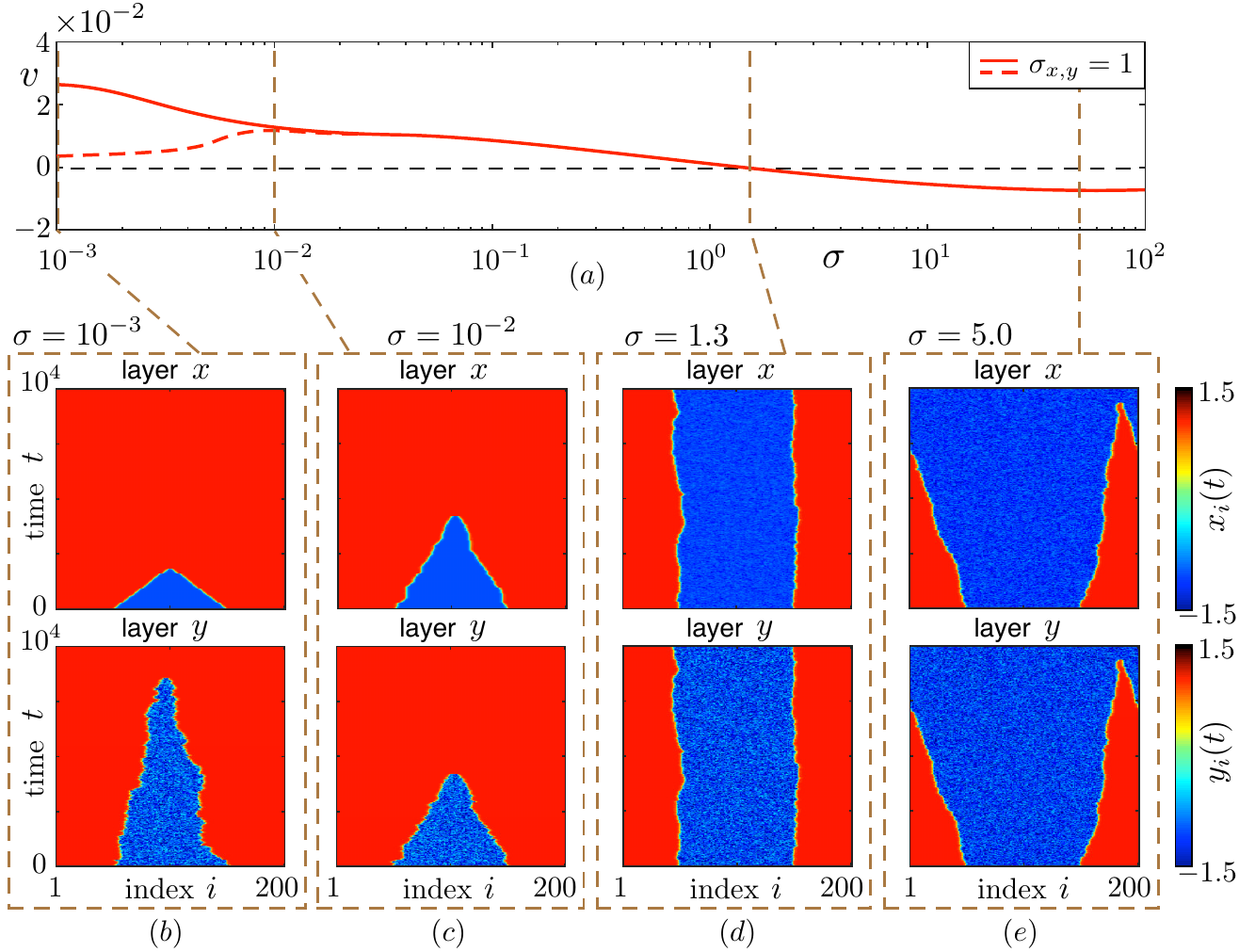}
\caption{Multiplexing-based control of front propagation (coarsening) in a two-layer multiplex network (Eqs. (\ref{two-layers}) in the presence of noise: (a) Dependence of the mean front propagation speed on the multiplexing strength in layer $x$ (solid curves) and $y$ (dashed curves) for intra-layer coupling $\sigma_{x,y}=1$; (b)-(e) Spatio-temporal dynamics of layer $x$ (upper panels) and $y$ (lower panes) for fixed  intra-layer coupling ($\sigma_x=\sigma_y=1$) and increasing multiplexing strength: $\sigma=10^{-3}$ (panels (b)), $\sigma=10^{-2}$ (panels (c)), $\sigma=1.3$ (panels (d)), $\sigma=5.0$ (panels (e)). Other parameters: $a_x=1.0$, $b_x=0.9$, $a_y=1.0$, $b_y=0.9$, $D=0.14$.}
\label{fig3}
\end{figure*}  
%%%%%%%%%%%%%%%%%%%%%%%%%%%%%%%%%%%%%%%%%%

\subsection{Deterministic model}
First, we consider the deterministic system (see Eqs. (\ref{two-layers}) at $D=0$). To reveal and visualize the action of multiplexing, we study two interacting layers. The first layer consists of asymmetric ($a_x=1.0$, $b_x=0.9$) oscillators, while the second-layer elements are symmetric ($a_y=1.0$, $b_y=1.0$). The intra-layer coupling strength is fixed. The initial conditions for the first layer are the same as in Sec. 2: $x_i(t=0)=-b_x$ for $i\in [50:150]$ and $x_i(t=0)=a_x$ elsewhere. The initial conditions for the second layer are similar: $y_i(t=0)=-b_y$ for $i\in [50:150]$ and $y_i(t=0)=a_y$. When we vary the multiplexing strength, the numerical simulations carried out for each $\sigma$ value start from the same initial conditions mentioned above.

To quantitatively describe the front propagation, we introduce the propagation speed of the left front moving from the left to the right: $v=\Delta i/\Delta t$. Here, $\Delta i$ represents the number of oscillators where the state $a_x$ or $a_y$ spreads over time $\Delta t$ (illustrated in Fig. \ref{fig2} (c)). This approach is based on the characterization of the front propagation in reaction-diffusion systems modeled in discretized space, where the front propagation distance is measured as a number of passed nodes multiplied by the space step, $s=\Delta i \times$step, and the corresponding speed takes the form $v=s/\Delta t$. 

%%%%%%%%%%%%%%%%%%fig4%%%%%%%%%%%%%%%%%%%%%%
\begin{figure*}[t]
\centering
\includegraphics[width=0.9\textwidth]{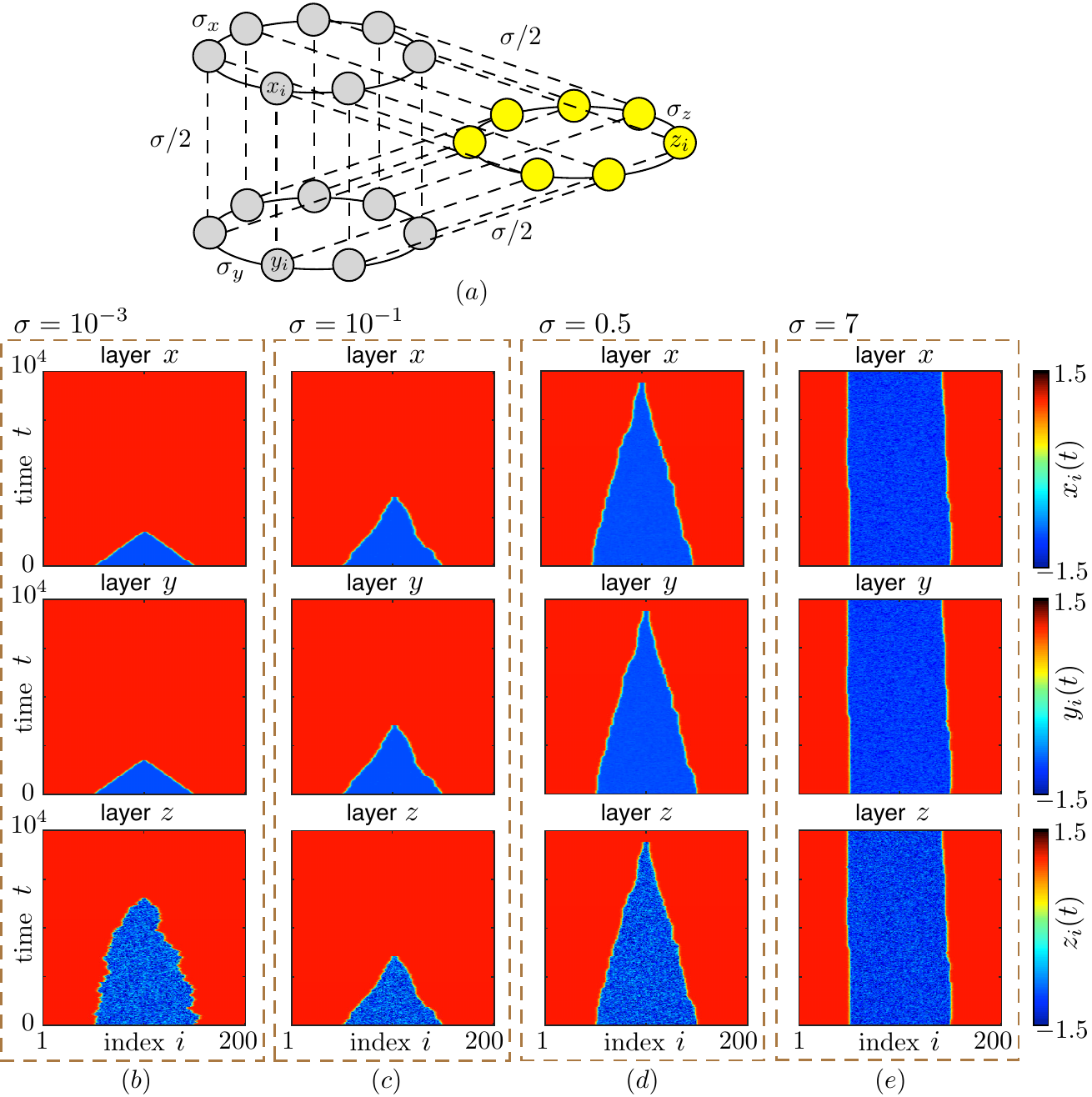}
\caption{Multiplexing-based control of front propagation (coarsening) in a three-layer multiplex network Eqs. (\ref{three-layers}) in the presence of noise: (a) Schematic representation of a three-layer network; (b)-(e) Spatio-temporal dynamics of layers $x$, $y$, $z$ for fixed  intra-layer coupling ($\sigma_x=\sigma_y=\sigma_z=1$) and increasing multiplexing strength: $\sigma=10^{-3}$ (panels (b)), $\sigma=10^{-1}$ (panels (c)), $\sigma=0.5$ (panels (d)), $\sigma=7.0$ (panels (e)). Other parameters: $a_x=a_y=a_z=1.0$, $b_x=b_y=b_z=0.9$, $D=0.14$.}
\label{fig4}
\end{figure*}  
%%%%%%%%%%%%%%%%%%%%%%%%%%%%%%%%%%%%%%%%%%

The increase of the multiplexing strength in the model Eqs. (\ref{two-layers}) causes transformations illustrated in Fig. \ref{fig2} (b) as the dependence of front propagation speed in the first (solid curves in Fig. \ref{fig2} (b)) and second (dashed curves in Fig. \ref{fig2} (b)) layer for two different values of the intra-layer coupling strength: $\sigma_x=\sigma_y=1$ and $\sigma_x=\sigma_y=2$. In both cases, the front propagation speed in layers $x$ and $y$ tends to the same value when the strength of multiplexing is increased. However, increasing the multiplexing strength allows to stop the front propagation in the layer of coupled asymmetric oscillators for weak intra-layer coupling (for instance, see the red curves in Fig. \ref{fig2} (b) corresponding to $\sigma_x=\sigma_y=1$), while in the ring of symmetric elements the wavefronts remain to be motionless. If the intra-layer coupling strength is large enough, then the front propagation speeds approach each other and tend to identical non-zero values (see the blue curves in Fig. \ref{fig2} (b) corresponding to $\sigma_x=\sigma_y=2$). This effect is illustrated in Fig. \ref{fig2} (c)-(e). For weak coupling between the layers their dynamics is quite similar to the front propagation in the isolated layer: one observes the front propagation in the first layer $x$, while the fronts in the second layer do not propagate [Fig. \ref{fig2} (c)]. Increasing $\sigma$, one induces the front propagation in the second layer and slows down this motion in the first layer [Fig. \ref{fig2} (d)]. Then the front propagation speeds tend to each other, which finally results in identical spatio-temporal diagrams for both layers: the fronts propagate with the same speed for sufficiently strong interaction between the layers [Fig. \ref{fig2} (e)]. After the front propagation speeds became identical, they do not vary for further increasing multiplexing strength.

\subsection{Stochastic model}
Next, we consider two layers where each oscillator is asymmetric: Eqs. (\ref{two-layers}) for $a_x=a_y=1.0$, $b_x=b_y=0.9$ and intra-layer coupling $\sigma_x=\sigma_y=1$. The initial conditions are the same as before. In contrast to Sec. 3.2, the intensity of noise in the second layer is non-zero $D=0.14$ which corresponds to the front propagation stabilization in the isolated layer. Thus, layers $x$ and $y$ evolve as depicted in Fig. \ref{fig1}~(b, c) in the absence of inter-layer coupling. The transformation of wavefront propagation caused by increasing the multiplexing strength is illustrated in Fig. \ref{fig3} (a) as the dependence of the mean front propagation speed on the multiplexing strength. The mean front propagation speed is obtained by numerical simulations of the model Eqs. (\ref{two-layers}) repeated 10 times starting from the same initial conditions. First, the mean speeds approach each other [Fig. \ref{fig3} (b)] and tend to the same value [Fig. \ref{fig3} (c)] as in the deterministic case. However, further increase of the inter-layer coupling strength results in subsequent changes in the mean propagation speed, which is the main difference between front propagation control in deterministic and stochastic models. In this way, one can achieve stabilized fronts (where $v_{x,y}=0$) in both layers at an appropriate level of multiplexing $\sigma\approx 1.3$ [Fig. \ref{fig3} (d)]. After passing through the multiplexing strength $\sigma\approx 1.3$, the front propagation in both layers is reversed [Fig. \ref{fig3} (e)]. Thus, adjusting the coupling between the layers, one can slow down, stabilize and reverse the wavefront propagation. 

It is important to note that zero wavefront propagation speed achieved in the multilayer network Eqs. (\ref{two-layers}) is characterized by less deviations of instantaneous front position in comparison with the single-layer model Eq. (\ref{single-layer}) (compare motions of fronts in Fig. \ref{fig1} (c) and Fig. \ref{fig3} (d)). It means that multiplexing can be applied for the minimization of front fluctuations. This constructive effect can be enhanced by increasing the number of interacting layers. To demonstrate this, let us consider a three-layer multiplex network schematically depicted in Fig.~\ref{fig4}~(a):
\begin{equation}
\label{three-layers}
\begin{array}{l}
\dfrac{dx_{i}}{dt}=-x_{i}(x_{i}-a_x)(x_{i}+b_x)\\
+\dfrac{\sigma_{x}}{2}\sum\limits^{i+1}_{j=i-1}(x_{j}-x_{i})+\dfrac{\sigma}{2}(y_{i}-x_{i})+\dfrac{\sigma}{2}(z_{i}-x_{i}), \\
\dfrac{dy_{i}}{dt}=-y_{i}(y_{i}-a_y)(y_{i}+b_y)\\
+\dfrac{\sigma_{y}}{2}\sum\limits^{i+1}_{j=i-1}(y_{j}-y_{i})+\dfrac{\sigma}{2}(x_{i}-y_{i})+\dfrac{\sigma}{2}(z_{i}-y_{i}),\\
\dfrac{dz_{i}}{dt}=-z_{i}(z_{i}-a_z)(z_{i}+b_z+\sqrt{2D}n_i(t))\\
+\dfrac{\sigma_{z}}{2}\sum\limits^{i+1}_{j=i-1}(z_{j}-z_{i})+\dfrac{\sigma}{2}(x_{i}-z_{i})+\dfrac{\sigma}{2}(y_{i}-z_{i}).
\end{array}
\end{equation}
Suppose that the layers represent rings of locally-coupled asymmetric bistable oscillators, $a_x=a_y=a_z=1.0$ and  $b_x=b_y=b_z=0.9$ where the asymmetry of the third layer is compensated by multiplicative noise, $D=0.14$ and the coupling strength within the layers is fixed, $\sigma_x=\sigma_y=\sigma_z=1.0$. As illustrated in Fig.~\ref{fig4}~(b)-(e), increasing the multiplexing strength gives rise to the effects which are similar to those observed in the two-layer network Eqs. (\ref{two-layers}). However, the achieved stabilization in the three-layer network is characterized by lower deviations of wavefronts in comparison with the two-layer network (compare Fig. \ref{fig3} (d) and Fig. \ref{fig4} (e)). The results of numerical simulations indicate almost negligible wavefront deviations for further increasing the number of interacting layers.

\section{\label{sec:conclusion}Conclusion}
It has been established that multiplexing can be applied for inducing and controlling wavefront propagation in multilayer networks of bistable oscillators. In particular, multiplexing allows to reduce the wavefront propagation to a common regime where the front propagation speed in all the interacting layers tends to the same value. This value is fixed in the deterministic system, but varies in the stochastic model with increasing the multiplexing strength. This provides for stopping (stabilization) and reversing the front propagation in all the interacting layers. It is important to note that the stabilization of wavefront propagation in stochastic multilayer networks is characterized by reduced deviations of the instantaneous wavefront position in comparison with the single-layer topology. 

Besides the possibility for control of the wavefront propagation speed by varying the multiplexing strength, the constructive role of multiplexing is manifested by the minimization of wavefront deviations. The minimization is enhanced by increasing the number of interacting layers. As reported recently \cite{semenov2022}, multiplexing provides for enhancement of stochastic resonance and this impact is also reinforced for the increasing amount of interacting layers. Thus, one can conclude that the ability to enhance the constructive influence of multiplexing on stochastic dynamics of bistable multilayer networks by increasing the number of interacting layers has a general character.

The present work is the first step towards a detailed study of the multiplexing-based wavefront propagation control and raises a number of questions. In particular, the theoretical background of the observed phenomena remains to be understood. These and other questions are issues for further investigations. 

\section*{Declaration of Competing Interest}
The authors declare that they have no known competing financial interests or personal relationships that could have appeared to influence the work reported in this paper.

\section*{Data Availability}
The data that support the findings of this study are available from the corresponding author upon reasonable request.

\section*{Acknowledgements}
We acknowledge support by the Deutsche Forschungsgemeinschaft (DFG, German Research Foundation) -- Projektnummer -- 163436311-SFB-910. Results of numerical simulations presented in Sec. 3.1 and 3.2 are obtained by V.V.S. in the framework Russian Science Foundation (Grant No. 22-72-00038).

%%% Loading bibliography style file
%%\bibliographystyle{model1-num-names}
%%%%%%%%%%%%%%%%%%%%%%%\bibliographystyle{cas-model2-names}
%
%% Loading bibliography database
%%%%%%%%%%%%%%%%%%%%%%%\bibliography{bibliography}

\end{document}